\documentclass[prb,nofootinbib,twocolumn,superscriptaddress]{revtex4} 
\usepackage{graphicx}% Include figure files
\usepackage{dcolumn}% Align table columns on decimal point
\usepackage{bm}% bold math
\usepackage{threeparttable}
\usepackage{times}
\usepackage{mathptmx}
\usepackage{lscape}
\usepackage{natbib}
\usepackage{amsmath}
\usepackage{braket}

\def\degree{\kern-.2em\r{}\kern-.3em}

\begin{document}

%\preprint{APS/123-QED} 

\title{ First-principles study on segregation of ternary additions for MoSi$_2$/Mo$_5$Si$_3$ interface  }

\author{Koretaka Yuge}
\affiliation{
Department of Materials Science and Engineering,  Kyoto University, Sakyo, Kyoto 606-8501, Japan\\
}%

\begin{abstract}
We investigate segregation behavior of additive elements $M$ (= Ni, Cr)  at the C11$_{\textrm{b}}$/D8$_{\textrm{m}}$ interface for MoSi$_2$-Mo$_5$Si$_3$ alloys, based on first-principles calculation. We first find energetically stable interface structure with interface energy of 0.08 eV/\AA$^2$. Based on the stable interface, segregation energy for additive elements is calculated for individual atomic layer, which is applied to Monte Carlo statistical simulation under grand-canonical ensemble to quantitatively predict interface segregation profile. We find that our simulation successfully capture the characteristics in measured segregation tendency of (i) Similarity in strong segregation at interface both for Ni and Cr compared with bulk composition, and (ii) stronger segregation for Ni than for Cr, which can be mainly attributed to differences in calculated segregation energy. The present results indicate that measured segregated interface for MoSi$_2$-Mo$_5$Si$_3$ alloys can be thermodynamically stable. 
\end{abstract}
 %\pacs{81.30.-t \sep 64.70.Kb \sep 64.75.+g }% PACS,  the Physics and Astronomy

\maketitle

\section{Introduction} 
%\begin{itemize}
 For super-high temperature structural materials, refractory transition-metal silicides are amply investigated so far, which can effectively improve the performance of such as gas turbine engine in power generation systems. 
 Particularly, MoSi$_{2}$ with C11$_{\textrm{b}}$ structure has been highly focused on, since it exhibits high melting temperature, high oxidation resistance, and low-temperature plastic deformability.\cite{Mo01,Mo02,Mo03,Mo04,Mo05,Mo06,Mo07} 
 However, in order to apply MoSi$_{2}$ to industrial applications, modification is still required since it exhibits poor fracture toughness, and poor creep strength at high temperatures.\cite{Mox01,Mox02} 
 In order to modify these drawbacks, extensive works have been performed to form duplex composite with MoSi$_{2}$.\cite{Nb03,Nb04,Nb05} 
 Mo$_{5}$Si$_{3}$ with D8$_{\textrm{m}}$ structure can be a promising candidate since MoSi$_{2}$/Mo$_{5}$Si$_{3}$ composite exhibits high eutectic temperature with script lamellar microstructures: 
 The recent study confirms that the eutectic MoSi$_{2}$/Mo$_{5}$Si$_{3}$ composite can significantly modify the creep strength of the 
MoSi$_{2}$ much effectively than other MoSi$_{2}$-based composits previously reported, while low-temperature fracture toughness should be still further modified.\cite{MM1, MM2, MM3} 

 For modification of the fracture toughness, introducing additional elements to effectively change the interface cohesion has been performed: They find that additional elements with low solubility in both MoSi$_{2}$ and Mo$_5$Si$_3$, including Ni and Co, exhibit pronounced segregation to MoSi$_2$/Mo$_5$Si$_3$ interface, and also find that introducing these elements successfully refine the microscopic structure of script lamellar.\cite{kishida2014} 
 Although these results strongly indicate that interface segregation of additional elements can effectively control its microscopic structure, it has not been theoretically confirmed (i) energetically stable contact in atomic scale between MoSi$_2$ and Mo$_5$Si$_3$ without additional elements, or (ii) whether the segregated interfaces are thermodynamically stable.  
 With these considerations, the present study address these two important points by using cluster expansion\cite{CE1,CE2} technique based on first-principles calculations. The previous experimental study confirm that the interface mainly composed of the so-called "terrace" and "ledge" part, where we here focus on the stable contact for terrace without additional elements, and then quantitatively estimate temperature dependence of segregation profile near interface for additional elements of Ni and Co. 

%\end{itemize}

\section{Methodology}
\subsection{Stable contact between C11$_{\textrm{b}}$ and D8$_{\textrm{m}}$}
%\begin{itemize}
 Let us first describe how to find energetically most stable contact, where crystal structures of MoSi$_2$ with C11$_{\rm b}$ and Mo$_5$Si$_3$ with D8$_{\textrm{m}}$ structure are shown in Fig.~1. 
\begin{figure}[h]
\begin{center}
\includegraphics[width=0.74\linewidth]{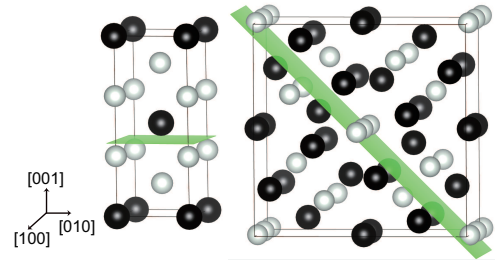}
\caption{ Crystal structure of MoSi$_2$ (C11$_\textrm{b}$)  and Mo$_5$Si$_3$ (D8$_{\textrm{m}}$) with their lattice plane of (001) and (110) described by green planes, where dark and bright spheres respectively represents Mo and Si atom. }
\label{fig:1}
\end{center}
\end{figure}

 The previous experimental study reveals that the terrace part of the interface corresponding to the contact between C11$_{\rm b}$ (001) and D8$_{\textrm{m}}$ (110) plane, where C11$_{\rm b}$ [110] and [1$\overline1$0] are respectively parallel to D8$_{\textrm{m}}$ [1$\overline1$0] and [001] direction. 
 From Fig.~1, it is clearly seen that there can be multiple number of possible symmetry-nonequivalent contacts. 
 In the first-principles calculation, we therefore construct all possible symmetry-nonequivalent contacts in order to find energetically most stable interface without additional elements. 
 Since considered lattice misfit between $d_{110}$ for MoSi$_2$ with C11$_{\textrm{b}}$ and $d_{330}$ for Mo$_5$Si$_3$ with D8$_{\textrm{m}}$ is sufficiently small (i.e., $\sim 0.3$\%), we here construct coherent interface structure where the lattice parameter is kept fixed at that for MoSi$_2$ with C11$_{\textrm{b}}$. 
 The interface slab for first-principles calculation is composed of 18-layer C11$_{\textrm{b}}$ (001), 24-layer D8$_{\textrm{m}}$ (110) plane and 19 \AA$ $ vacuum region, which leads to 116 Mo atoms and 120 Si atoms in the slab. We consider the energetic stability for the interface slabs based on interface energy.$\gamma$\cite{yuge-seg02} 
 We performed the first-principles calculation using a DFT code, the Vienna ab initio simulation package (VASP)\cite{vasp1,vasp2}  based on the projector augmented wave method,\cite{paw,paw2} to obtain total energy for the interface slabs. Generalized gradient approximation Perdew-Burke-Ernzerhof (GGA-PBE)\cite{pbe} was employed to treat the exchange-correlation functional. Plane-wave cutoff energy of 400 eV was used throughout the calculations. Geometry optimization was performed until the residual forces became less than 1 meV/\AA. Brillouin-zone integration was performed on the basis of the Monkhorst-Pack scheme\cite{MP} with a $2\times 4\times 1$ k-point mesh. 
%\end{itemize}
\subsection{Interface segregation of additional elements}
%\begin{itemize}
 We describe here our model to address interface segregation behavior of additional elements. For segregation, we employ energetically most stable interface obtained by the above procedure. In order to compare our first-principle calculation with previous experimental reports, we choose two additional elements of Ni and Co, which both exhibits strong interface segregation.\cite{kishida2014} 
 Then we define corresponding interface segregation energy for additional element $M$ ($M$=Ni or Co) as
\begin{eqnarray}
\label{eq:seg}
\Delta E_\textrm{seg}^{M}\left( \Lambda \right) = E_\textrm{seg}^{M}\left( \Lambda \right) - E_\textrm{seg}^{\textrm{Mo5Si3}},
\end{eqnarray} 
where  $E_\textrm{seg}^{\textrm{Mo5Si3}}$ denotes DFT energy of the interface slab where one 
Mo atom in the middlemost layer in D8$_{\textrm{m}}$ region is replaced by $M$, and $E_\textrm{seg}^{M}\left( \Lambda \right)$ denotes DFT energy where one Mo atom in the other layer $\Lambda$ in D8$_{\textrm{m}}$ or C11$_{\textrm{b}}$ is replaced by $M$. 
% Note that $E_\textrm{seg}^{M}\left( \Lambda \right)$ corresponds lowest energy interface slab among possible intralayer-replacement of Mo by $M$. 
 We employ the same calculation condition for first-principles as that for finding stable interface structure described above. 
%\end{itemize}

\section{Resutls and Discussions}
%\begin{itemize}
 We first show in Fig.~2 interface energy $\gamma$ for possible contact between C11$_\textrm{b}$ (001) and D8$_{\textrm{m}}$ (110) plane (left-hand figure), and interface structure having lowest energy (right-hand figure). We can clearly see that interface energy for possible contacts all have positive interface energy, and they are typically around 0.1-0.3 eV/\AA$^2$. These values of interface energy is slightly higher than those for interface between C11$_{\textrm{b}}$ and C40 of MoSi$_2$ confirmd by our previous DFT study,\cite{yuge-seg02} which naturally comes from the deviation in geometric differences: For the latter case, main difference comes from their stacking sequence, while in the present system, this does not hold true. 
\begin{figure}[h]
\begin{center}
\includegraphics[width=0.92\linewidth]{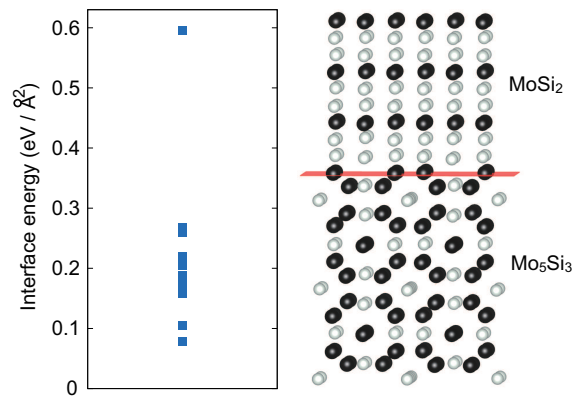}
\caption{Left: Interface energy for possible contacts between MoSi$_2$ C11$_{\textrm{b}}$ (001) and Mo$_5$Si$_3$ D8$_{\textrm{m}}$ (110) plane. Right: Interface structure having lowest interface energy, where dark and bright spheres denotes Mo and Si atom, respectively. }
\label{fig:2}
\end{center}
\end{figure}
 From Fig.~2, we can qualitatively see that energetically stable interface structure appears to have coherent contact with each other, having interface energy of 0.08 eV/\AA$^2$. 
 
Next, using the most stable interface shown in Fig.~2, we estimated segregation energy for additive elements $M$ ($M$=Ni or Co) to address energetically favorable site for additional elements. Figure~3 shows segregation energy for the three additive elements in terms of distance from interface. 
\begin{figure}[h]
\begin{center}
\includegraphics[width=0.88\linewidth]{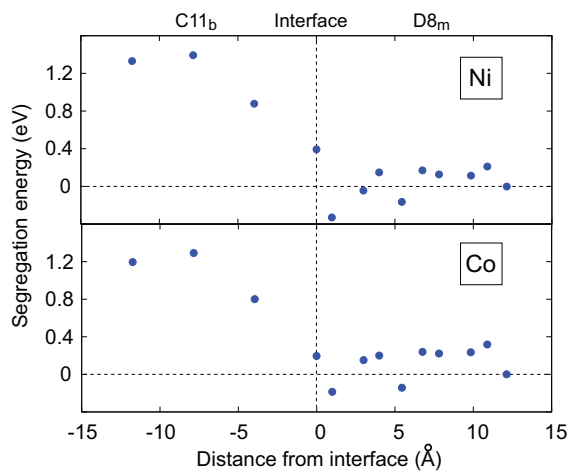}
\caption{ Segregation energy for Ni and Co in terms of distance from interface. }
\label{fig: }
\end{center}
\end{figure}
 We can clearly see that segregation energy shows qualitatively similar tendencies: (i) They all exhibit minimum value at distance from interface of around 1 \AA, indicating that Ni and Co are all expected to exhibit strong segregation to interface, and (ii) segregation energy for MoSi$_2$ region is around 1 eV higher than that for Mo$_5$Si$_3$ region: These (i) and (ii) both qualitatively agree with the previous experimental result\cite{kishida2014} where composition of Ni or Co for Mo$_5$Si$_3$ region is around 0.5 \% higher than that for MoSi$_2$ region. 
 Therefore, the present approach based on segregation energy for additive elements defined in Eq.~(\ref{eq:seg}) can reasonably capture the characteristics of the measured interface segregation. 

 To quantitatively determine the temperature-dependence of segregation profile at the interface, we combine cluster expansion (CE)  technique with Monte Carlo (MC) statistical simulation to include statistical ensemble. 
We have demonstrated the predictive power of the present combination of CE and MC based on DFT calculation in the segregation profile for alloy nanoparticles and surfaces.\cite{yg1,yg2,yg3,yg4} 
Briefly, CE provides orthonormal expansion of internal energy in terms of atomic configurations, where their basis functions are described by a pseudospin variable, $\sigma_i$, taking +1 (-1) when $i$ site is occupied by Mo (Ni or Co) atom. In the CE, configurational energy is given by
\begin{eqnarray}
E\left( \vec{\sigma} \right) = \sum_{\alpha} V_{\alpha} \Braket{ \prod_{i\in \alpha}\sigma_i },
\end{eqnarray}
where $\Braket{\quad}$ represents linear average over all lattice points, and coefficient, $V_{\alpha}$, is called effective cluster interaction (ECI) for cluster $\alpha$ consisting of lattice points. To apply the CE expansion, we employed the following relationship between segregation energy in Eq.~(\ref{eq:seg}) and ECI:\cite{yuge-seg,yuge-seg02}
\begin{eqnarray}
\label{eq:V}
V_{1}^{\left( \Lambda \right)} - V_{1}^{\textrm{Mo5Si3}} = \frac{\Delta E_{\textrm{seg}}^{M}\left( \Lambda \right)}{2}, 
\end{eqnarray}
where $V_{1}^{\left( \Lambda \right)}$ denotes ECI for point cluster at interface layer $\Lambda$. Using the above relationship, segregation profile for Ni and Co can be quantitatively estimated by applying the ECIs to MC simulation under grand-canonical ensemble. 
In the MC simulation, an interface slab of $12 \times 12$ in-plane expansion of the interface with 28 layer is used in the calculation. We confirm that the size of the used MC simulation box is sufficient for ensemble average: We search chemical potential giving dilute Ni or Co composition of 0.5 \% in D8$_{\textrm{m}}$ bulk region, since the measured Ni or Co  composition for this region is around this value.\cite{kishida2014}

 Under these conditions, we show in Fig.~4 the resultant interface segregation profile for Ni and Co at $T=1673$ K, in terms of distance from interface. 
We can clearly see the strong segregation at interface both for Ni and Co, which agrees with previous experimental result. 
\begin{figure}[h]
\begin{center}
\includegraphics[width=0.94\linewidth]{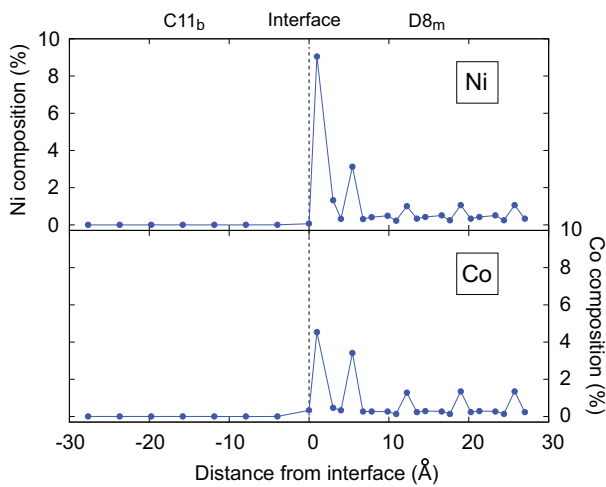}
\caption{Segregation profile at $T=1673$ K for Ni and Co in terms of distance from interface, predicted by Monte Carlo statistical simulation. }
\label{fig:4}
\end{center}
\end{figure}
 Furthermore, Ni composition at interface is much higher (more than twice) than Co composition at interface. These results successfully capture the relative interface segregation for Ni and Co by previous measurement.\cite{kishida2014} 
 Meanwhile, we find quantitative difference in absolute magnitude of composition at interface: Our simulation predicts Ni and Co composition at interface of 9.1 and 4.5 \%, while previous experiment reports lower value of $\sim 3$ and $\sim 1$ \%, respectively. 
 We believe this deviation between our theoretical and previous experimental results mainly comes from the fact that while our simulation can predict composition for additive elements at individual atomic layer (order of \AA), experimental composition is averaged over much wider region of order of nano meter: Considering these facts, our thermodynamic simulation successfully predict interface segregation for Ni and Co, indicating that segregated interface between C11$_{\textrm{b}}$ (001) and D8$_{\textrm{m}}$ (110) plane is thermodynamically stable, which can be reasonablly attributed to differences in segregation energy shown in Fig.~3 for energetically most stable interface. 
%\end{itemize}

\section{Conclusions}
Segregation behavior of additive elements of Ni and Cr for MoSi$_2$/Mo$_5$Si$_3$ interface is quantitatively investigated based on first-principles calculation. We focus on segregation profile for Interface structure having lowest interface energy, where corresponding segregation energy is considered for Monte Carlo statistical simulation under grand-canonical ensemble. We find that strong interface segregation and its magnitude relationship for Ni and Cr can be reasonablly attributed to differences in segregation energy of individual atomic layer for most stable interface: 
These facts indicate that previously measured segregated interface can be thermodynamically stable.

\section*{Acknowledgement}
This research was supported by Advanced Low Carbon Technology Research and Development Program of the Japan Science and Technology Agency (JST).

\end{document}